# Rescue Network: Using UAVs (drones) in Earthquake Crisis Management

**Masoud Hayeri Khyavi**

m.hayery@itrc.ac.ir

**Abstract:** Earthquake is one of the natural disasters which cannot be either controlled or predicted absolutely. Since preventing earthquake is impossible, preventing its damages is also difficult. Unfortunately, after each earthquake and its financial and life losses, the initial panic of the people results in the second wave of accidents and damages. Inrush of confused people to escape the cities, streets and houses is a great problem. Apart from training in seismic areas which is very important, considering security arrangements and observing security principles in construction, instructing the people is also important. Other than searching for and rescuing the people who are trapped under detrital or are in danger, those who thieve the damaged area is another important issue after each earthquake. Thus, a solution is proposed to use modern technology to reduce threats of natural disasters including earthquake.

Today, UAVs are being used in natural disasters and accidents. To this end and considering the ever-increasing developments of network technologies and communication including IoT and cloud, an efficient design is presented which increases rescue factor of live creatures in natural disasters that can be used to rescue human lives and prevent subsequent outcomes after a few seconds. In this study, focus is on time of occurrence of earthquake and after earthquake.

## Proposed Solution

In this scheme, some platforms are considered for the flight of drones considering the specified areas.

The proposed solution is a Mapping of IoT and cloud networks (Figure 1) and Botnet (Figure 2). In other words, bot networks can be used to program and control the operation and a combination of sensors, actuators, communications and protocols are used in this structure

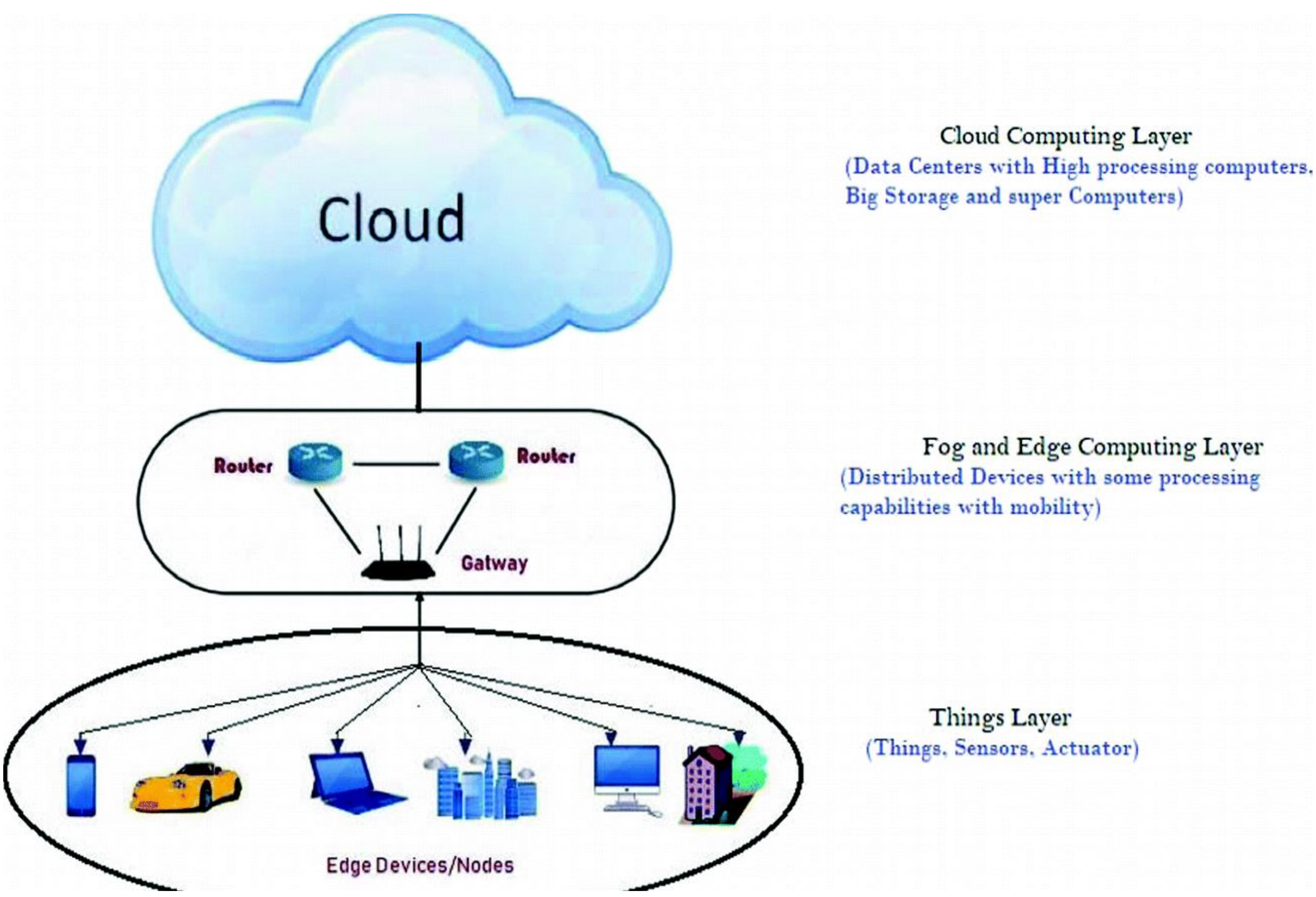

**Figure1. Fog and edge computing with cloud architecture in IoT environment** [1]

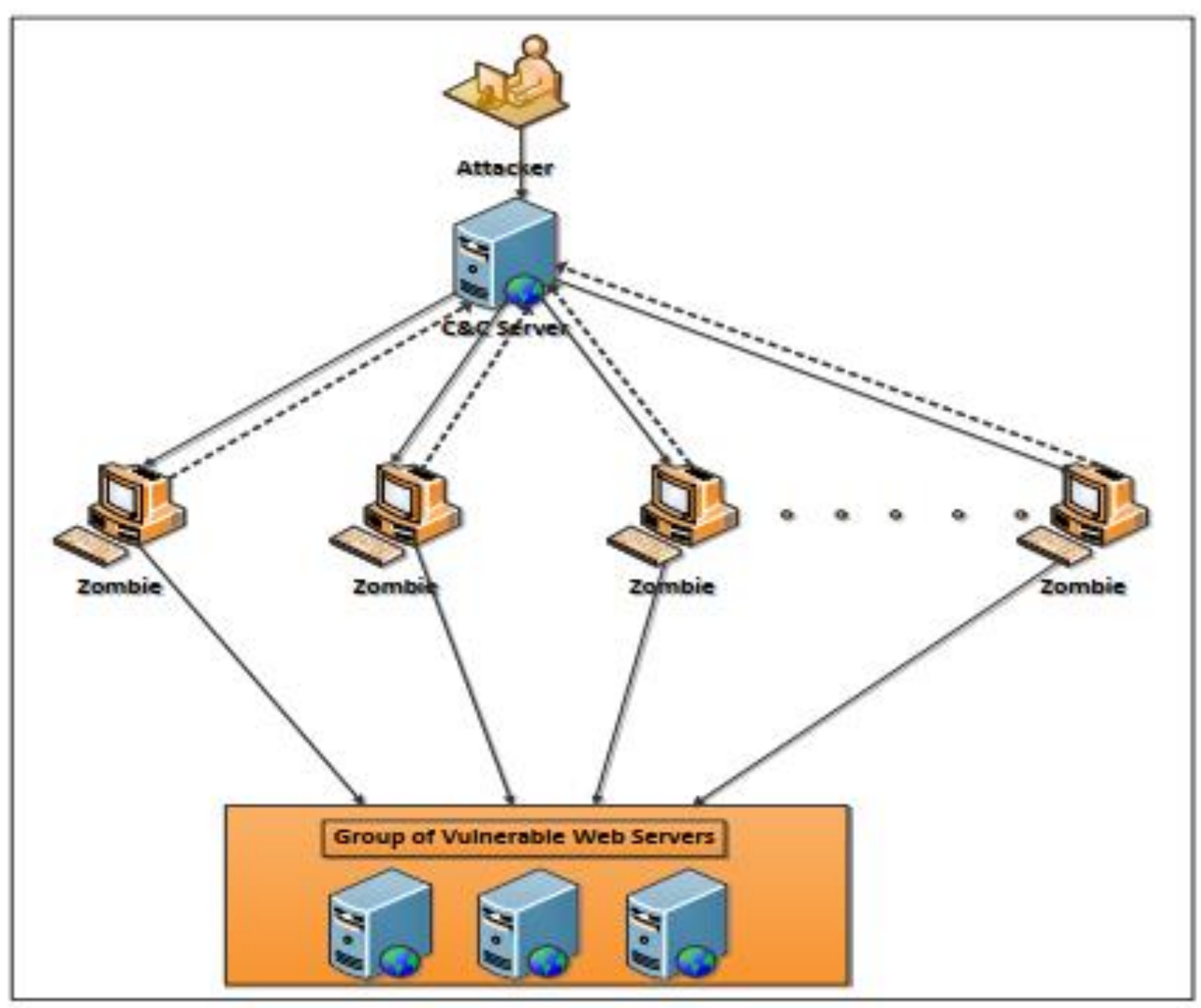

**Figure 2. HTTP-based Botnet Structure** [2]

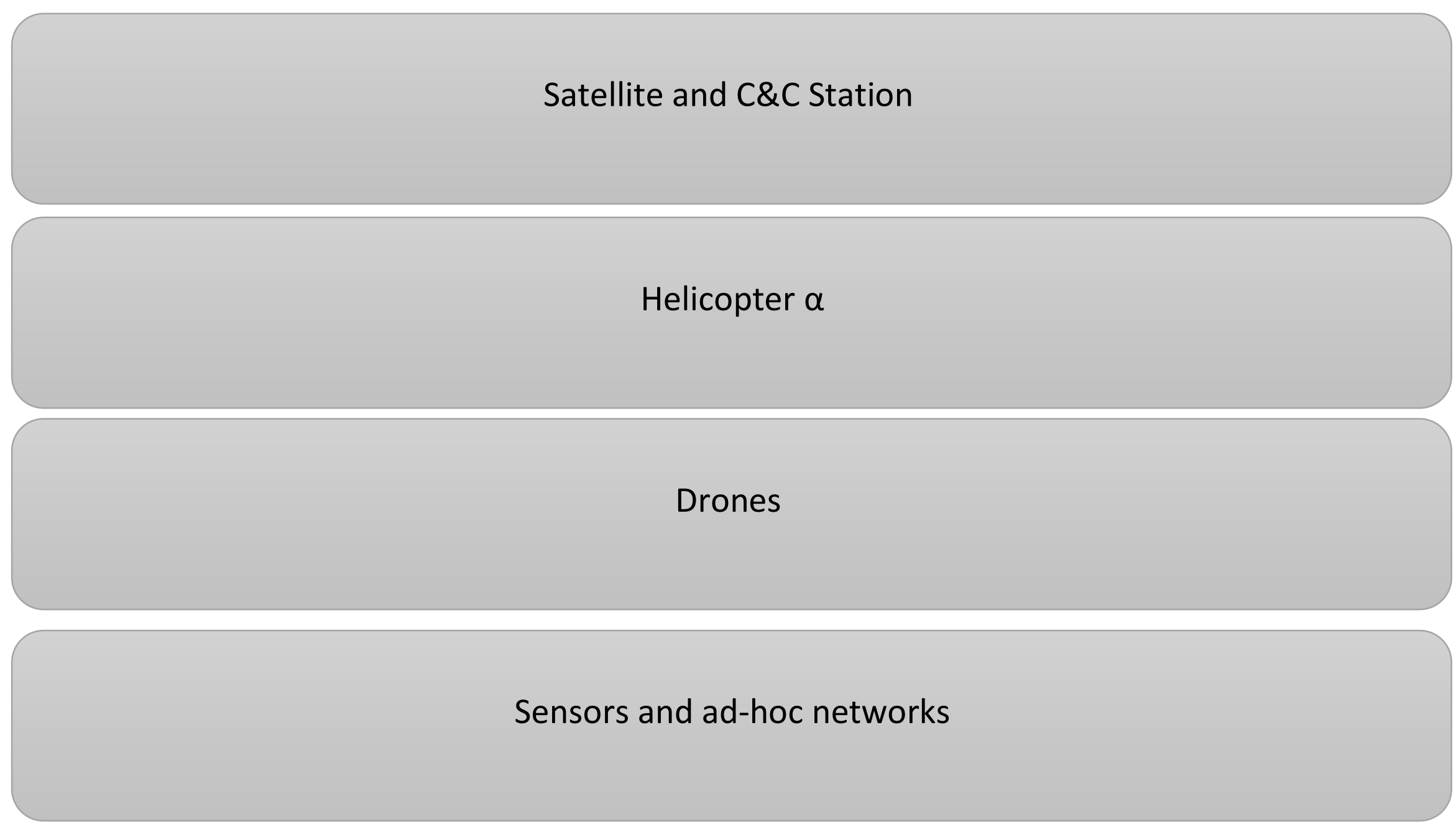

**Figure 3 . The proposed layered structure of the rescue network**

## A. Sensors

Sensors or actuators or specific smart devices are distributed in the city considering position of the faults. According to Figure 1, sensors are located in the host layer. The city is divided into high risk, medium risk and low risk areas considering strength of the fault (Figure 4):

- In the high-risk area (H), devices with high reliability in terms of physics and communication are located. These devices are connected to specific smart drones, point by point.
- In the medium-risk area (M), devices with high reliability in terms of communication are located.
- In the low-risk area (L), devices with high reliability in terms of accuracy and computation are located.

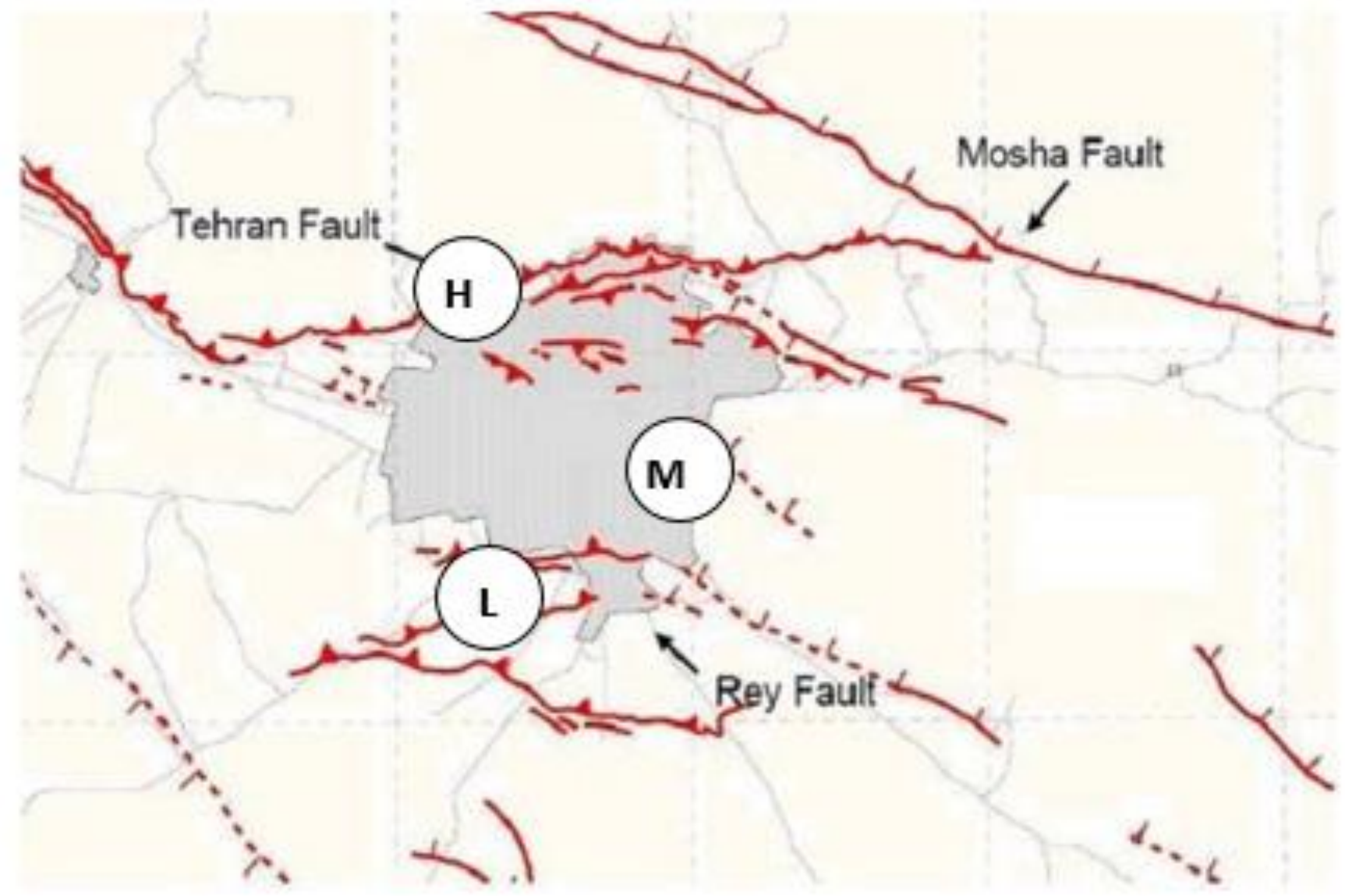

**Figure 4. Segmentation based on fault strength (Tehran)**

Performance is as follows (Figure 5):

## 1. At the time of earthquake

- First step: upon receiving vibrations or abnormal oscillations (compared based on a predefined threshold), these devices transmit the data to the upper layer (edge or fog) which has a computational server with limited processing and the smart drones.
- Second Step: The drones immediately take action and take off the ground and go to the predefined areas. In the second layer or edge, alarms are in the yellow level and transmit the result of computations and the data received from drones to helicopter α and the ground stations.
- In helicopter α, initial computations are performed and the instructions are activated and the drones get ready to perform the predefined instructions.
- Information is transmitted simultaneously from helicopter α to the crisis management center and the satellite.
- The satellite takes action because the probability that connections are disrupted is very high in severe earthquakes.
- Upon receiving information from the satellite and helicopter α, the crisis management center makes new decisions and manages the crisis.

## B. Helicopter

Helicopters take two roles of α and β and make three operations.

- They are used for surveillance (helicopter β)
- They act as a backup for the drones which are disabled and there is no alternative drone to replace (helicopter β)
- Helicopter α acts as a server to receive, compute and transmit information to the drones, satellites and rescue centers. In other words, it acts as a coordinator guided and controlled by the management center.

**C. Drones**

- Drones are prepared and equipped to take action. Drones can be manufactured for a specific purpose.
- Drones have monitoring, communication and computation devices.
- Specific stations are considered in the city for their deployment.
- These stations are connected to the seismic centers and alarm (before earthquake) and react fast (after earthquake).

Drones are ready to fly and take action in the following situations:

- Upon receiving alarms from the geophysics and seismic centers
- Upon sensing severe vibrations based on what is defined for them (via sensors)

Drones performs as follows:

- Upon sensing sever earthquakes, they automatically, take off the ground and deploy in the zoned areas.
- They connect with helicopter and other adjacent drones.
- They analyze their coverage area.
- If a problem occurs, they inform the nursing drone.
- The nursing drone has to monitor if the drones are healthy and if a problem occurs, it acts based on the predefined instruction and another drone takes this responsibility automatically (if there is no drone, helicopter β takes this responsibility)
- Each drone has detection (which finds live creatures) sensors.
- If a problem occurs for a drone in a specific zone, another drone takes its responsibility automatically

Drones at points A, B, C and D communicate with the followings based on the defined facilities via helicopter α:

- Communication satellites
- Ground stations
- Police
- Ground rescue

**Communications**

Communications are either point by points, wireless or satellite

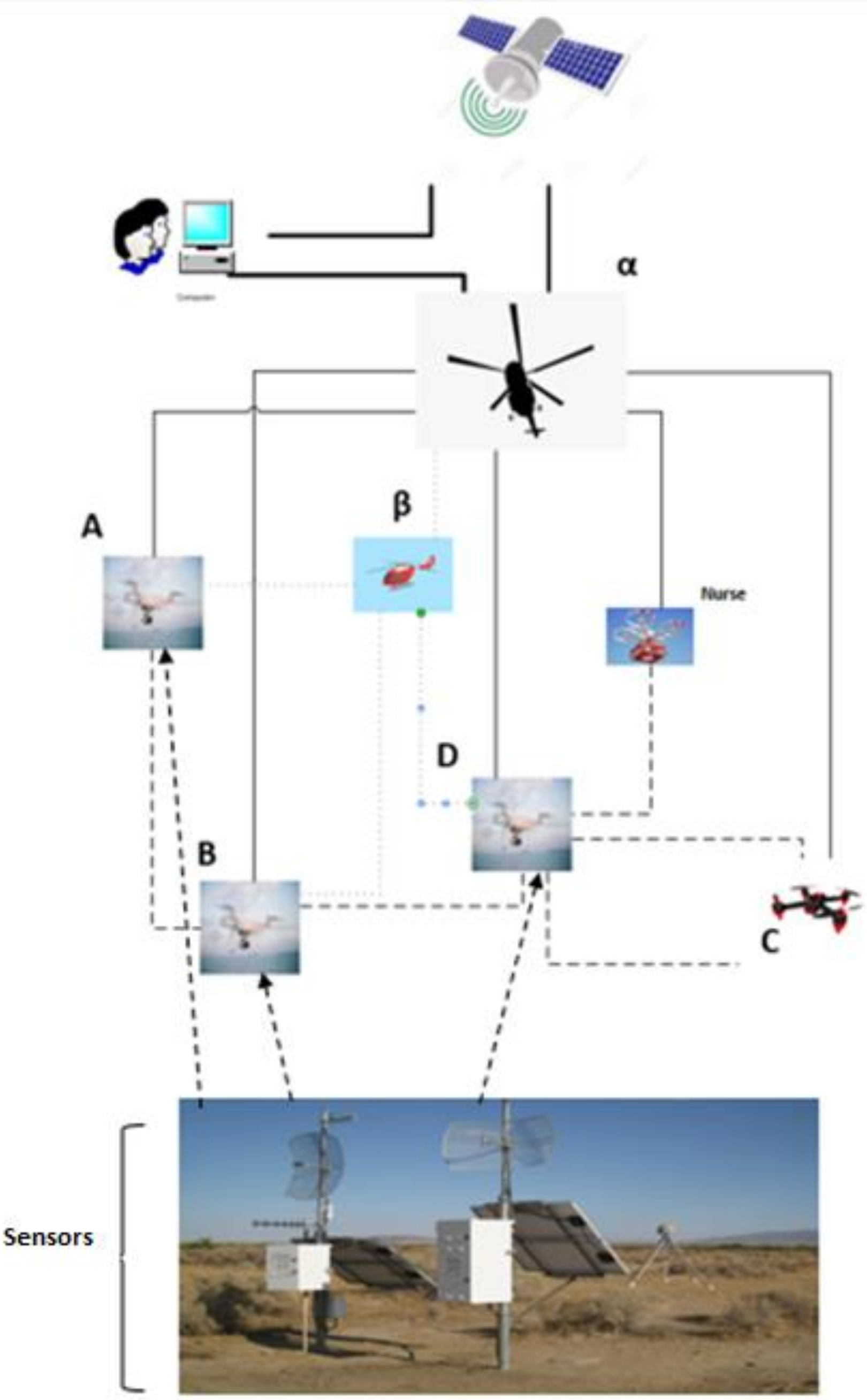

**Figure 5 . Scheme of equipment and communications**

## 1. After Earthquake

Using the gathered information, preserving and updating communications, the rescue team can do their best to perform their responsibilities.

Drones can detect the traffic caused by people who are escaping and inform the rescue team about congested areas so that they can manage the traffic. If a route is closed or drones can land, the route is specified and the exit route is changed to a secure direction.

Secure areas other than those specified in the crisis preparation phase are detected and specified by the drones. Drones can guide the population towards secure areas (via light at nights). Drones reports the prior areas which need to be rescued and secured. On the other hand, drones can use the modern technology to detect the places in which live creatures are trapped and the rescue team can rescue the creatures trapped under detrital faster (using the access route information).

## Conclusion

Earthquake is one of the natural disasters which cannot be either controlled or predicted absolutely. Since preventing earthquake is impossible, preventing its damages is also difficult. Unfortunately, after each earthquake and its financial and life losses, the initial panic of the people results in the second wave of accidents and damages. Inrush of confused people to escape the cities, streets and houses is a great problem. Apart from training in seismic areas which is very important and considering security arrangements and observing security principles in construction, instructing the people is also important. Along with finding and rescuing the people who are trapped under detrital or in a dangerous, those who thieve the damaged area is another important issue after each earthquake. Thus, a solution is proposed to use modern technology to reduce threats of natural disasters including earthquake. This scheme increases rescue factor of live creatures in natural disasters and it can be used to rescue humans and prevent subsequent outcomes after a few seconds. In addition, this scheme can be used to design and implement other schemes for other disasters.